\documentclass[prl,twocolumn,floatfix,tighten,bibtex,letterpaper]{revtex4}
\usepackage{graphicx}
\usepackage{bm}
\usepackage{amssymb}
\usepackage{color}
\usepackage{epsfig}
\usepackage{subfigure}

\newcommand{\be}{\begin{equation}}
\newcommand{\ee}{\end{equation}}
\def \be{\begin{equation}}
\def \ee{\end{equation}}
\def \ba{\begin{array}}
\def \ea{\end{array}}
\def \bea{\begin{eqnarray}}
\def \eea{\end{eqnarray}}

\begin{document}

\title{Unconventional Josephson signatures of Majorana bound states}
\author{Liang Jiang$^1$, David Pekker$^1$, Jason Alicea$^2$, Gil Refael$^1$,
Yuval Oreg,$^3$and Felix von Oppen$^4$}
\affiliation{$^1$Department of Physics, California Institute of Technology, Pasadena,
California 91125, USA}
\affiliation{$^2$ Department of Physics and Astronomy, University of California, Irvine,
CA 92697, USA}
\affiliation{$^3$Department of Condensed Matter Physics, Weizmann Institute of Science,
Rehovot, 76100, Israel }
\affiliation{$^4$Dahlem Center for Complex Quantum Systems and Fachbereich Physik, Freie
Universit\"at Berlin, 14195 Berlin, Germany}

\begin{abstract}
A junction between two topological superconductors containing a pair
of Majorana fermions exhibits a `fractional' Josephson effect, $4\pi$
periodic in the superconductors' phase difference. An additional fractional
Josephson effect, however, arises when the Majoranas are spatially separated
by a superconducting barrier. This new term gives rise to a set of Shapiro
steps which are essentially absent without Majorana modes and therefore
provides a unique signature for these exotic states.
\end{abstract}

\maketitle

Majorana fermions comprise the simplest and likely most experimentally
accessible non-Abelian anyon. An unambiguous demonstration of their
non-Abelian exchange statistics would be a great triumph for condensed
matter physics, as this phenomenon reflects one of the most spectacular
manifestations of emergence. Furthermore, non-Abelian excitations provide
the foundation behind topologically protected quantum computation \cite%
{Kitaev-TQC,TQCreview}, with Majorana fermions playing a crucial role in
prototype devices \cite%
{Nayak-TQC,Bonderson-measure-only,Hassler,JiangKanePreskill,QuantumBus}. In
the solid-state context, Majorana modes were originally perceived as
zero-energy states bound to vortices in $p$-wave superconductors \cite%
{Volovik-pwave}, and therefore are also associated with quasi-particles in
the Moore-Read state \cite{ReadGreen}. More recent proposals employ
topological insulators \cite{FuKane-edge, FuKane3d,CookFranz}, {half-metals
in proximity to superconductors \cite{BrouwerDuckheim, Lee-NCS-SC,
Zhang-half-metals}}, as well as spin-orbit-coupled quantum wells \cite%
{Sau2DEG,Alicea2DEG} and nanowires \cite%
{Lutchyn10,Oreg10,JiangKitagawa,HoleDopedWire} to stabilize these elusive
particles. Signatures of Majorana fermions appear in tunneling spectra and
noise \cite{BolechDemler,MIRAR}, and more strikingly through interference
effects \cite{ShtengelIF,SternIF}.

Josephson effects provide yet another important experimental signature of
Majorana fermions. Kitaev first predicted that a pair of Majoranas fused
across a junction formed by two topological superconducting wires generates
a Josephson current \cite{Kitaev01}
\begin{equation}
I=\frac{e}{\hbar}J_M\sin\left(\frac{\phi_{\ell}-\phi_r}{2}\right),
\label{J1}
\end{equation}
which exhibits a remarkable $4\pi$ periodicity in the superconducting phase
difference $\phi_{\ell}-\phi_r$ between the left and right wires. In stark
contrast to ordinary Josephson currents, this contribution reflects
tunneling of \emph{half} of a Cooper pair across the junction. Such a
`fractional' Josephson effect was later established in other systems
supporting Majorana modes \cite%
{FuKane-edge,FuKane3d,Lutchyn10,Oreg10,Ioselevich}, and in direct junctions
between p-wave superconductors \cite{SenguptaYakovenko}. In this manuscript
we demonstrate that two topological superconductors bridged by an ordinary
superconductor with phase $\phi_m$ generically support a second kind of
unconventional Josephson effect with an associated current
\begin{equation}
I^{\prime }=\frac{e}{\hbar}J_Z\sin\left(\frac{\phi_{\ell}+\phi_r}{2}%
-\phi_m\right),  \label{J2}
\end{equation}
in the right or left superconductors, and twice that in the middle. This
contribution arises solely from the fusion of spatially-separated Majoranas
across the junction, and represents processes whereby a Cooper pair in the
middle region splinters, with half entering the left and half entering the
right topological superconductor. We will derive this emergent term in 1d
Majorana-supporting systems, and propose several ways of measuring its
effects.

This novel Josephson coupling is derived most simply in a 1d Kitaev chain.
Consider a junction with Hamiltonian $H = H_\ell + H_r + \delta H$, where
the left/right superconductors are described by 
$p$-wave-paired spinless fermions $c_{\alpha,x}$ ($\alpha=\ell,\,r$) hopping
on an $N$-site chain \cite{Kitaev01},
\begin{equation}
H_{\alpha} = - \sum_{x = 1}^{N-1}(t c_{\alpha,x}^\dagger c_{\alpha,x+1} +
\Delta e^{i\phi_{\alpha}} c_{\alpha,x} c_{\alpha,x+1}+ h.c.).  \nonumber
\label{Hkitaev}
\end{equation}
Eq.\ (\ref{Hkitaev}) adiabatically connects to realistic Majorana-supporting
quantum wire Hamiltonians \cite{Lutchyn10,Oreg10,Alicea10}, and therefore
describes their universal properties as well. Following Kitaev, we express
the spinless fermions in terms of two Majorana operators via $c_{\alpha,x} =
\frac{1}{2}e^{-i\frac{\phi_{\alpha}}{2}}(\gamma^{\alpha}_{B,x} + i
\gamma^{\alpha}_{A,x})$. When $t = \Delta$,
Eq.\ (\ref{Hkitaev}) maps onto a dimerized Majorana chain: $H_{\alpha} = -it
\sum_{x = 1}^{N-1}\gamma^{\alpha}_{B,x}\gamma^{\alpha}_{A,x+1}$. The
explicit absence of $\gamma_{A,1}^{\alpha}$ and $\gamma_{B,N}^{\alpha}$ in
the Hamiltonians indicates the presence of zero-energy Majorana modes
localized at the ends of each superconductor in the junction.

Let us now couple the two superconductors through
\begin{equation}
\delta H = -t_m (c_{\ell,N}^\dagger c_{r,1} + h.c.) -
\Delta_m(e^{i\phi_m}c_{\ell,N} c_{r,1} + h.c.),
\end{equation}
where the two terms describe tunneling and Cooper pairing across the
junction. These couplings combine the zero-energy Majorana modes residing at
the junction into a finite-energy Andreev bound state.
Focusing on these zero-energy modes, one can write $c_{\ell,N} \rightarrow
\frac{1}{2}e^{-i\phi_{\ell}/2}\gamma_{B,N}^\ell$ and $c_{r,1} \rightarrow i
\frac{1}{2}e^{-i\phi_r/2}\gamma_{A,1}^r$, and define an ordinary fermion
operator $f^\dagger = \frac{1}{2}(\gamma_{B,N}^\ell + i \gamma_{A,1}^r)$; $%
\delta H$ then becomes
\begin{eqnarray}
\delta H &\rightarrow& (2f^\dagger f -1)\{J_M\cos[(\phi_\ell-\phi_r)/2]
\nonumber \\
&+& J_Z \cos[(\phi_\ell + \phi_r)/2-\phi_m]\}.  \label{hm}
\end{eqnarray}
with $J_M=\frac{t_m}{2}$ and $J_Z=\frac{\Delta_m}{2}$. Since the current in
region $s$ is given by $\frac{2e}{\hbar}\frac{\partial \langle {\delta H}%
\rangle}{\partial \phi_s}$, the fermion tunneling $t_m$ gives rise to the
fractional Josephson effect of Eq.\ (\ref{J1}), while pairing $\Delta_m$
across the junction produces the Josephson current in Eq.\ (\ref{J2}). Note
that the sign of either current is dictated by the occupation number for the
$f$ fermion, and hence can be used as a readout method for qubit states
encoded by the Majoranas \cite{FuKane3d,Alicea10}.

A more quantitative understanding is obtained by considering more realistic
models. Let us consider Majoranas localized on a
topological insulator edge in proximity to a superconductor and subjected to
a magnetic field \cite{FuKane-edge}; a very similar analysis applies to
quantum wires. In the Nambu spinor basis $\Psi^T=(\psi_{\uparrow},\psi_{%
\downarrow},\psi_{\downarrow}^{\dagger},-\psi_{\uparrow}^{\dagger})$, the
Bogoliubov-de Gennes Hamiltonian for this system is
\begin{equation}
{\mathcal{H}}= v \hat{p} \sigma^z \tau^z -\mu\tau^z +\Delta
\left(\cos\phi\tau^x-\sin\phi\tau^y\right) +B \sigma^x,  \label{hwire}
\end{equation}
with $v$ the edge-state velocity, $\hat{p}$ the momentum, $B$ the Zeeman
energy, and $\sigma^a$ and $\tau^a$ Pauli matrices acting in the spin and
particle-hole sectors, respectively. We allow the chemical potential $\mu$,
pairing amplitude $\Delta$, and superconducting phase $\phi$, to vary
spatially.

\begin{figure}[tbp]
\includegraphics[width=9cm]{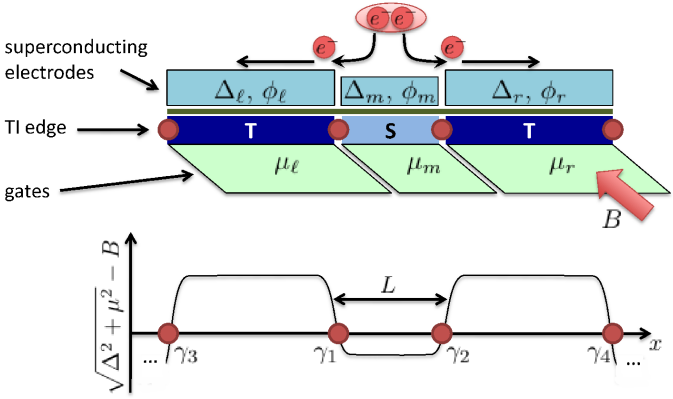}
\caption{Topological insulator edge subjected to a magnetic field $B$ and
sandwiched by gates and superconducting electrodes. Majorana modes (red
circles) localize at domain walls where the gap $E_{gap}=|\protect\sqrt{%
\protect\mu^2+\Delta^2}-B|$ vanishes and the argument in the absolute value
changes sign. When the middle region is a trivial superconductor (S), and
the sides form a topological phase (T), the novel $J_Z$ term in Eq. (\protect
\ref{hm}), with current $\propto \sin(\frac{\protect\phi_{\ell}+\protect\phi%
_r}{2}-\protect\phi_m )$, accompanies the usual the fractional Josephson
effect. This splits a Cooper pair in the middle electrode into two single
electrons, injected via the two Majorana states in each topological segment.
The same effect appears in spin-orbit-coupled wires in a T-S-T
configuration. }
\label{3domains}
\end{figure}

Majorana states arise at interfaces between topological (T) and trivial (S)
regions of the edge \cite{FuKane-edge}. With $\mu$, $\Delta$, and $\phi$
uniform the quasi-particle gap is $E_{gap}=|B-\sqrt{\Delta^2+\mu^2}|$. When $%
\sqrt{\Delta^2 + \mu^2} > B$ the edge is gapped by proximity-induced
superconductivity and forms a topological phase closely related to that of
Kitaev's model described above \cite{FuKane-edge}. In the trivial phase $%
\sqrt{\Delta^2 + \mu^2} < B$, and the magnetic field dominates the gap. We
will study the T-S-T domain sequence of Fig. \ref{3domains}, which localizes
Majoranas $\gamma_1$ at $x=0$ and $\gamma_2$ at $x=L$. Each of the three
regions, $\ell,r,m$, couples to a superconductor imparting proximity
strength $\Delta_{\ell/m/r}$ and phase $\phi_{\ell/m/r}$, and has a chemical
potential $\mu_{\ell/m/r}$ controlled by separate gates. (The main
difference in the quantum wire case is that there creating the T-S-T domain
structure needed to observe the unconventional Josephson effects discussed
here requires the reversed criteria: $\sqrt{\Delta^2 + \mu^2} < B$ in the
outer regions and $\sqrt{\Delta^2 + \mu^2} > B$ in the middle region.)

The Majorana-related Josephson effects result from hybridization between $%
\gamma _{1}$ and $\gamma _{2}$. When $\gamma _{1}$ and $\gamma _{2}$ are far
apart ($L\rightarrow \infty $), they constitute exact zero-energy modes, and
their wave functions decay exponentially in region $s=\ell ,m,r$ with \emph{%
two} characteristic lengths:
\begin{equation}
\lambda _{s\pm }=\frac{v}{|\Delta _{s}\pm \sqrt{B^{2}-\mu _{s}^{2}}|}
\label{lambda}
\end{equation}%
(we assume $\mu _{s}<B$). For finite $L$, however, $\gamma _{1,2}$ combine
into a finite-energy state with creation operator $f^{\dagger }=\frac{1}{2}%
(\gamma _{1}+i\gamma _{2})$. Roughly, each Majorana perceives the interface
localizing the other Majorana as a perturbation, yielding a hybridization
which is suppressed as a weighted sum of two decaying exponentials. This
hybridization is again described by Eq.\ (\ref{hm}), with $J_{M/Z}=\frac{1}{2%
}\left( J_{+}e^{-L/\lambda _{m+}}\pm J_{-}e^{-L/\lambda _{m-}}\right) $. An
explicit calculation (see supp. material) for the symmetric setup, $\mu
_{\ell }=\mu _{r}\equiv \mu $, $\Delta _{\ell }=\Delta _{r}\equiv \Delta
>\Delta _{m}$ and $\mu _{m}=0$ yields 
\begin{equation}
J_{+}=J_{-}\approx \frac{2\Delta }{\frac{\Delta (B+\Delta )+\mu ^{2}}{\Delta
^{2}+\mu ^{2}-B^{2}}+\frac{B\Delta }{B^{2}-\Delta _{m}^{2}}}.  \label{pres}
\end{equation}%
When $\Delta _{m}=0$ and the middle region is normal---which is the setup
typically studied \cite{FuKane-edge,Lutchyn10}---$J_{Z}=0$ and hence only
the Josephson term in Eq.\ (\ref{J1}) appears. Turning on $\Delta _{m}\neq 0$
yields a nonzero $J_{Z}$, and the second Josephson term in Eq.\ (\ref{J2}).
Furthermore, since both $J_{Z}$ and $J_{M} $ are dominated by the slowest
decay length, they will generically be of the same order. For a quantitative
estimate, consider the parameters $\mu _{m}=0$, $\mu _{l,r}=E$, $\Delta
_{m}=E$, $\Delta _{l,r}=\sqrt{8}E$, $B=2E$ with energy scale $E=0.1$meV.
Assuming an edge velocity $v=10^{4}$m/s, for this choice we obtain $\lambda
_{m+}\approx 22$nm, $\lambda _{m-}\approx 66$nm, and $J_{\pm }\approx 0.12$%
meV. The effect then peaks at $L\approx 50$nm, which yields $J_{Z}\approx
0.022$ meV and $I_{Z}=\frac{e}{\hbar}J_Z\approx 5.3$nA.

These Josephson effects are simplest to understand conceptually when two
additional Majoranas, $\gamma_{3,4}$, straddle the T segments of the edge as
shown in Fig.\ \ref{3domains}. Let us define fermion operators $f_A = \frac{1%
}{2}(\gamma_1 + i \gamma_3)$ and $f_B = \frac{1}{2}(\gamma_2 + i \gamma_4)$,
and assume that the corresponding occupation numbers are initially $n_A = 1$
and $n_B = 0$. We will further employ a `perturbative' perspective and
promote the superconducting phases to quantum operators conjugate to the
Cooper pair number. One can then see that the Majorana operators in the term
$J_M (2f^\dagger f -1)\exp(i\frac{\phi_r-\phi_{\ell}}{2}) =
iJ_M\gamma_1\gamma_2\exp(i\frac{\phi_r-\phi_{\ell}}{2})$ hop a single
fermion across the S region, changing the state of the edge from $%
(n_A,n_B)=(1,0)$ to $(0,1)$. At the same time, the exponential passes a
charge $e$ from side to side. The combination of these processes makes the
term gauge invariant. The persistent superconducting current limit in this
case is apparent when we consider an additional tunneling event which
restores the parities of the T segments, moving a fermion back to the left
but with a Cooper pair hopping to the right. A similar perspective clarifies
the role of the $J_Z$ term---the Majoranas in $iJ_Z\gamma_1\gamma_2\exp\left[%
i\left(\frac{\phi_r+\phi_{\ell}}{2}-\phi_m\right)\right]$ also change the
parity of the two T segments, while the exponent removes a Cooper pair from
the middle region and adds charge $e$ to each T region (see Fig.\ \ref%
{3domains}).

\begin{figure}[tbp]
\includegraphics[width=9cm]{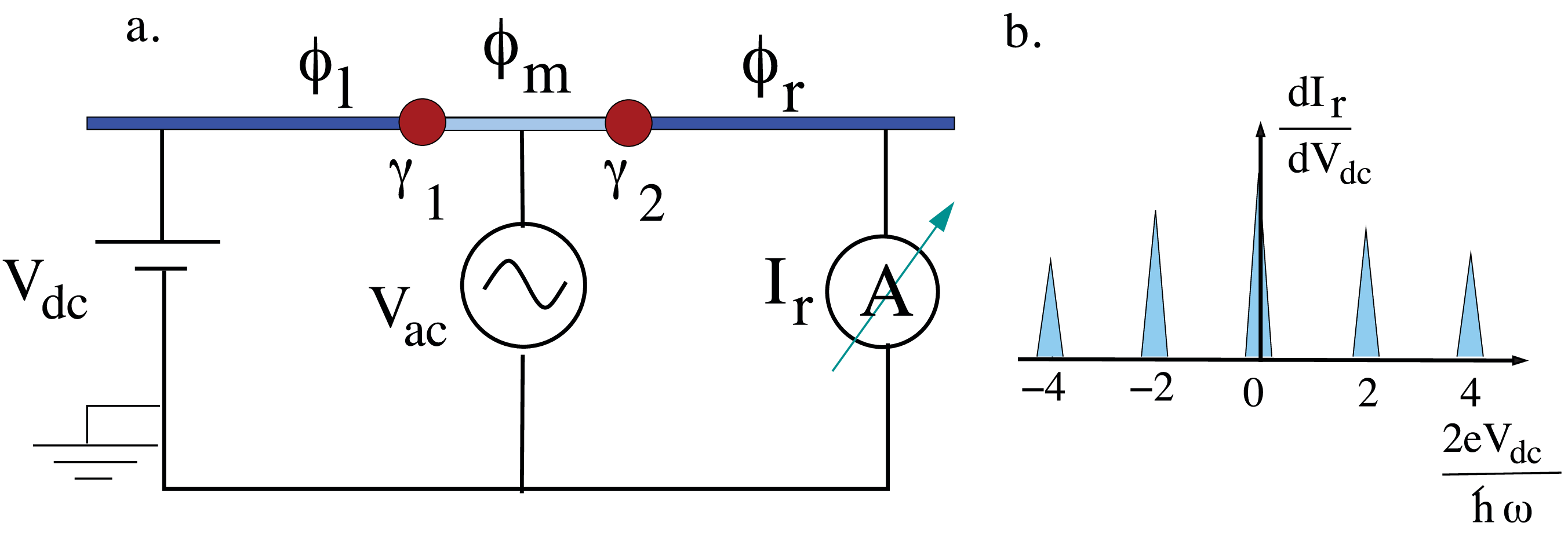}
\caption{Three-leg Shapiro-step measurement scheme. (a) We envision applying
a \emph{dc} voltage $V_{dc}$ to the left superconducting electrode and an
\emph{ac} voltage with angular frequency $\protect\omega$ in the left loop
(which we model as an \emph{ac} voltage applied to the middle electrode). A
measurement of the \emph{dc} current $I_r$ in the right electrode will then
reveal Shapiro steps stemming from the Majorana modes when the \emph{ac}
Josephson frequency $2eV_{dc}/\hbar$ equals an \emph{even} harmonic of $%
\protect\omega$. (b) Sketch of $dI_r/dV_{dc}$ indicating the predicted
Shapiro steps---note the crucial absence of odd-harmonic peaks, which would
appear in a conventional Shapiro-step measurement. }
\label{shapiro}
\end{figure}

Next, we discuss the crucial issue of measuring the new Josephson term in
Eq.\ (\ref{J2}).
The first and most direct possibility involves manipulating independently
the phase differences $\phi_{\ell}-\phi_m \equiv \Phi_L$ and $\phi_m-\phi_r
\equiv \Phi_R$, \emph{e.g.}, by inserting different fluxes in the two loops
in Fig.\ \ref{shapiro}a (ignoring the voltage sources in the figure). By
tuning $\Phi_L=-\Phi_R$ in a symmetric junction, one can probe the $J_Z$
Josephson term (driving current $J_Z\sin\Phi_L$ on the middle electrode)
while canceling the $J_M$ term. Such measurements, however, are highly
challenging---they require careful flux control; the Majorana-related
Josephson current must be disentangled from the conventional $2\pi$ periodic
contributions; and the measurement must be concluded before the parity of
the two Majoranas changes.


A potentially more promising measurement scheme relies on Shapiro steps. In
a regular Josephson junction, Shapiro steps arise from a combination of a
\emph{dc} voltage $V_{dc}$ and an \emph{ac} voltage $V_{ac}\sin\omega t$,
which together generate a current $I=I_J\sin\left[\phi_0+2eV_{dc}
t/\hbar-(2e V_{ac}/\hbar \omega)\cos\omega t\right]$. Naively, this current
averages to zero because of the constantly winding phase. This is not the
case, however, when $2eV_{dc}/\hbar=n\omega$ for some integer $n$---here a
\emph{dc} current component exists, producing a step in the $V$ vs.\ $I$
plot for the junction \cite{Shapiro,Tinkham}. For the fractional Josephson
term in Eq.\ (\ref{J1}), the $4\pi$ periodicity leads to Shapiro steps when $%
2eV_{dc}/\hbar=2n\omega$, corresponding to even Shapiro steps of a regular
Josephson junction. The halved periodicity, if established, could provide a
smoking-gun signature for Majorana modes. An inevitable conventional
Josephson current, however, `fills in' the missing steps, making it
difficult to disentangle these contributions \cite{SenguptaYakovenko}.

The following three-leg Shapiro-step measurement circumvents this problem
and targets the Josephson term of Eq.\ (\ref{J2}). As shown in Fig.\ \ref%
{shapiro}a, we envision a \emph{dc} voltage applied to the left leg so that $%
\phi_\ell=2eV_{dc}t/\hbar$, while an \emph{ac} voltage applied to the middle
leg sets $\phi_m=-(2eV_{ac}/\hbar\omega)\cos\omega t$. Since the new
Josephson term induces current in all three legs, a current measurement on
the right lead will find Shapiro steps emerging only when
\begin{equation}
2eV_{dc}/\hbar=2n\omega
\end{equation}
as illustrated in Fig.\ \ref{shapiro}b, without any odd-harmonic steps. This
non-local measurement is insensitive to any parasitic two-phase Josephson
terms, and therefore automatically eliminates most competing processes.
Furthermore, it bears the advantage of being a fast dynamic measurement
(since Josephson frequencies are typically in the GHz regime), which reduces
its sensitivity to temporal fluctuations of the Majorana-state occupations.

To verify the approximation methods used and to confirm the prominence of
the $J_Z$ term in the three-leg Shapiro measurement, we also numerically
analyzed the Josephson effects in a topological insulator edge. Figure~\ref%
{fig:JvsL} shows that our analytical results [\emph{e.g.}, Eq.\ (\ref{pres}%
)] indeed agree very well with the exact numerical calculation. We also
explored additional current contributions such as $\delta I
\sin(\phi_L+\phi_R-2\phi_m)$, which could obscure the Majorana signature by
producing unwanted odd-harmonic Shapiro steps. This term is independent of
the Majorana modes, and can instead arise from conventional Bogoliubov
states in the junction. In the limit of small pairing and tunneling over the
middle segments, such a term reflects a high-order process. Numerically, we
find that it is suppressed by at least an order of
magnitude compared to the Majorana $J_Z$ contribution in the regime where $%
J_Z$ is substantial, i.e., when $L$ is of order $\lambda_{m\pm}$.

\begin{figure}[tb]
\includegraphics[width=8.5cm]{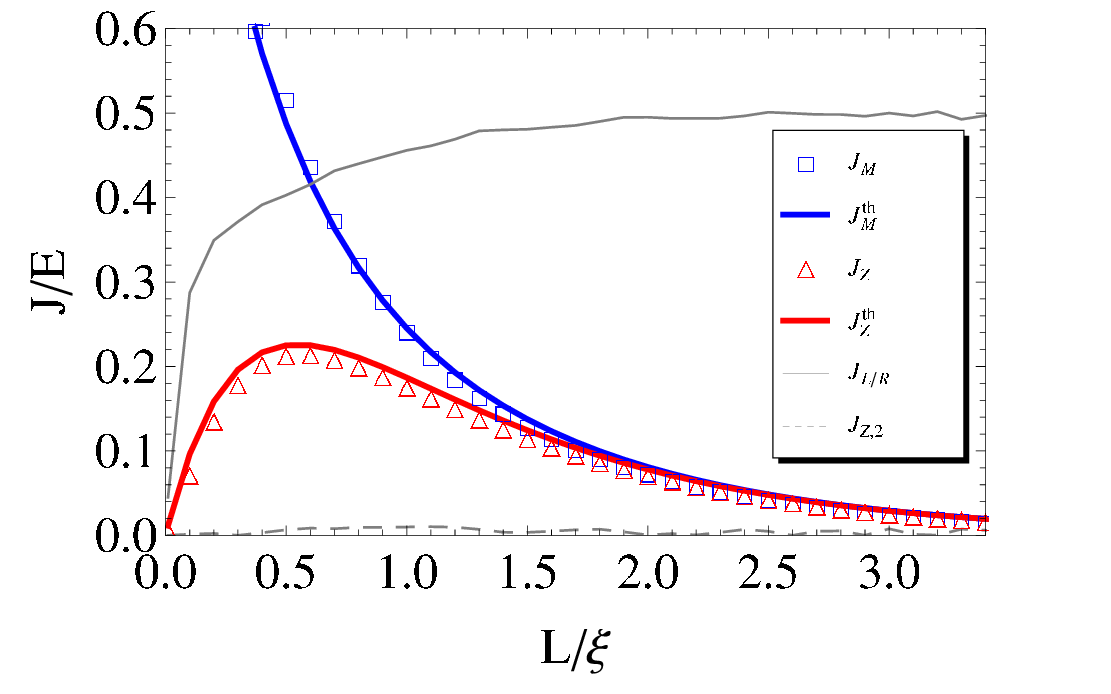}
\caption[fig:JvsL]{Numerically determined coefficients of conventional
Josephson couplings ($J_{L/R}$), Majorana-induced terms ($J_{M/Z}$), and
second harmonic of the $J_{Z}$ term ($J_{Z,2}$). Our analytical estimates of
$J_{M}^{th}$ and $J_{Z}^{th}$ agree well with numerics. The energy unit is $%
E $ and the length unit is $\protect\xi =v/E$. The parameters are $\protect%
\mu _{l,r}=E,\protect\mu _{m}=0$, $\Delta _{l,r}=\protect\sqrt{8}E$, $\Delta
_{m}=E$, and $B_{l,r}=B_{m}=2E$. The characteristic lengths are $\protect%
\lambda _{m+}=\protect\xi /3$ and $\protect\lambda _{m-}=\protect\xi $. For $%
E=0.1$meV and $v=10^{4}$m/s, the length unit is $\protect\xi =66$nm and the
maximum current is $I_{Z}=\frac{e}{\hbar}J_Z\approx 5.3$nA.}
\label{fig:JvsL}
\end{figure}

By considering the full edge spectrum (including the Andreev bound states
and continuum states exactly), we obtained the total Josephson energy of the
domain configuration in Fig.~\ref{3domains}:%
\begin{eqnarray}
E_{\mathrm{tot}} &\approx& J_{L}\cos (\phi _{\ell}-\phi _{m}) + J_{R}\cos
(\phi _{r}-\phi _{m})  \nonumber \\
&+& J_{M}\cos[(\phi _{\ell }-\phi _{r})/2] + J_Z \cos[(\phi_\ell +
\phi_r)/2-\phi_m]  \nonumber \\
&+& \sum_{n=2}^{\infty }J_{Z,n}\cos[n((\phi _{\ell }+\phi _{r})/2-\phi _{m})]
+\cdots
\end{eqnarray}
Here $J_{L/R}$ are conventional Josephson terms (to which the three-leg
measurement is insensitive), $J_{M/Z}$ are the Majorana-induced
contributions, and $J_{Z,n}$ denote the (unwanted) higher harmonics of the $%
J_Z$ term. As Fig.\ \ref{fig:JvsL} illustrates, $J_M$ dominates for $L\ll
\lambda_{m+}$, while for $\lambda_{m-}\gtrsim L \gtrsim \lambda_{m+}$ the $%
J_Z$ term becomes comparable, enabling the three-leg Shapiro-step
measurement. 
The higher harmonics $J_{Z,n}$ are at least an order of magnitude smaller
than $J_{Z}$ in this regime and can be neglected. For $L\gg \lambda_{m-}$
the Majorana signatures are strongly suppressed as expected.

In this manuscript, we explored a new Josephson effect that arises when a
pair of Majorana fermions fuse across a junction formed by two topological
superconductors separated by an ordinary superconductor. The Majoranas in
this setup enable Cooper pairs injected into the barrier superconductor to
`splinter' into the left and right legs of the junction---a process which
would ordinarily be prohibited at low energies. While Majorana modes can
also give rise to a novel fractional Josephson effect in T-normal-T
junctions, we argued that an important advantage of our setup is that here
one can more readily isolate the Majorana-mediated Josephson current through
Shapiro-step measurements. The experiments we proposed could provide a
relatively simple and unambiguous detection scheme for Majorana fermions,
and may also serve as a practical readout mechanism for qubit states encoded
by these particles.

It is a pleasure to thank M. P. A. Fisher, L. Glazman, J. Preskill, A. Kitaev, A. Stern, J. Meyer, K. Shtengel, C. Marcus, L. Kouwenhoven, and B. Halperin for useful discussions, and the Aspen Center for Physics for hospitality. We are also
grateful for support from BSF, SPP1285 (DFG), the Packard foundation, the
Sherman-Fairchild foundation, the Moore-Foundation funded CEQS, and the NSF
through grant DMR-1055522, and IQI grant number: PHY-0456720 and PHY-0803371.

\bibliographystyle{apsrev}
\bibliography{majorana}

\section{Supplementary material}

\section{Perturbative Calculation}

Let us pursue here a detailed calculation of the Josephson coupling across
the Majorana junction described in Fig. \ref{3domains}. We will first find
the wave functions of the Majorana states localized on each domain wall,
ignoring the existence of the other interface. We will denote these states
as $\left| L\right\rangle$ and $\left| R\right\rangle$. Next, we follow the
usual procedure for finding tight-binding states and Hamiltonians. We first
calculate the overlap matrix, $M_{\alpha\beta}=\left\langle
\alpha\right|\beta\rangle$ with $\alpha,\,\beta=L,\,R$, and the Hamiltonian
matrix within this subspace, $h_{\alpha\beta}=\left\langle \alpha\right|{%
\mathcal{H}}\left| \beta\right\rangle$. It is easy to see that the
approximate hybridization Hamiltonian is then given by
\begin{equation}
H_{maj}=M^{-1/2} h M^{-1/2}.  \label{tbm}
\end{equation}

\textbf{Single Majorana solution at $x=L$.} We first solve for the
zero-energy eigenstates of the Hamiltonian (\ref{hwire}) with parameters:
\begin{equation}
\begin{array}{c}
\Delta (x)=\Delta _{r}\Theta (x-L)+\Delta _{m}(1-\Theta (x-L)) \\
\mu (x)=\mu _{r}\Theta (x-L)+\mu _{m}(1-\Theta (x-L))%
\end{array}%
\end{equation}%
The solution has the same form on the two sides of the domain wall, but with
different parameters. We denote the side of the domain with the index $s$
being $s=r,\,m$ for right or middle. By squaring the Hamiltonian and looking
for momentum values yielding zero energy states, we find two imaginary
momenta on each side, which correspond to the spatial decay constants $%
\lambda _{s\pm }^{-1}$ given by Eq. (\ref{lambda}). The wave function
associated with each side of the domain is:
\begin{equation}
\left\vert r\right\rangle =\Psi _{s}(x)=\mathit{R}_{s+}\psi _{s+}e^{-\frac{%
|x-L|}{\lambda _{s+}}}+\mathit{R}_{s-}\psi _{s-}e^{-\frac{|x-L|}{\lambda
_{s-}}}.  \label{Psix}
\end{equation}%
with $\mathit{R}_{s\pm }$ being four complex numbers determining the
amplitude of the wave functions corresponding to the two decay lengths, and
with $\psi _{s\pm }$ being four, four-dimensional vectors, which when $\phi
_{s}=0$ are given by:
\begin{equation}
\psi _{m\pm }=\left(
\begin{array}{c}
1 \\
e^{-i\zeta _{m}} \\
\pm i \\
\mp ie^{-i\zeta _{m}}%
\end{array}%
\right) ,\,\,\psi _{r\pm }=\left(
\begin{array}{c}
1 \\
e^{\pm i\zeta _{r}} \\
i \\
-ie^{\pm i\zeta _{r}}%
\end{array}%
\right)
\end{equation}%
with $\exp (i\zeta _{s})=\frac{\mu _{s}+i\sqrt{B^{2}-\mu _{s}^{2}}}{B}$, for
$s=\ell ,\,r,\,m$. These solutions are the building blocks for each Majorana
state. In order to obtain what the wave function becomes when $\phi _{s}$
(the phases of the superconducting electrodes) deviate from zero, we can
apply the rotations: $\hat{U}_{\phi }=\exp \left( i\frac{\phi }{2}\tau
_{z}\right) $ such that:
\begin{equation}
\psi _{s\pm }^{(\phi _{s})}=\hat{U}_{\phi _{s}}\psi _{s\pm }.
\end{equation}%
Obtaining the Majorana solution follows from matching the boundary condition
of the solutions, and from them finding the coefficients $\mathit{R}_{s\pm }$%
.

To avoid the complicated expression that could arise in the most general
case of Majorana coupling, we concentrate on the case where $%
\Delta_r=\Delta_{\ell}>\Delta_m$ and $\mu_r=\mu_{\ell}=\mu$ and $\mu_m=0$.
This choice does not constitute a substantial loss of generality, and is
useful for grasping the results of our calculations. A straightforward but
rather tedious calculation leads to the following solution for the
amplitudes of the decaying waves of the right Majorana state under the above
assumptions:
\begin{equation}
\begin{array}{c}
\left(%
\begin{array}{c}
\mathit{R}_{m+} \\
\mathit{R}_{m-}%
\end{array}
\right)=\frac{2\sin\zeta_r}{1+ie^{-i\zeta_r}}\left(%
\begin{array}{c}
i\sin\left(\frac{\phi_r-\phi_m}{2}\right) \\
\cos\left(\frac{\phi_r-\phi_m}{2}\right)%
\end{array}
\right), \\
\left(
\begin{array}{c}
\mathit{R}_{r+} \\
\mathit{R}_{r-}%
\end{array}
\right)=\left(
\begin{array}{c}
-\frac{1+ie^{i\zeta_r}}{1+ie^{-i\zeta_r}} \\
1%
\end{array}
\right).%
\end{array}%
\end{equation}

By symmetry, we can infer the structure of the left Majorana, which is localized about $x=0$:
\begin{equation}
\left\vert \mathit{L}\right\rangle =\Psi _{s}^{(L)}(x)=\mathit{L}_{s+}\psi
_{s+}e^{{-\frac{|x|}{\lambda _{s+}}}}+\mathit{L}_{s-}\psi _{s-}e^{{-\frac{%
|x|}{\lambda _{s-}}}}.  \label{PsixL}
\end{equation}%
The amplitudes $\mathit{L}_{s\pm }$ also depend on the phases on the left
and middle segment of the wire in a similar way:
\begin{equation}
\begin{array}{c}
\left(
\begin{array}{c}
\mathit{L}_{m+} \\
\mathit{L}_{m-}%
\end{array}%
\right) =\frac{2\sin \zeta _{\ell }}{1-ie^{i\zeta _{\ell }}}\left(
\begin{array}{c}
-i\sin \left( \frac{\phi _{\ell }-\phi _{m}}{2}\right) \\
\cos \left( \frac{\phi _{\ell }-\phi _{m}}{2}\right)%
\end{array}%
\right) , \\
\left(
\begin{array}{c}
L_{\ell +} \\
L_{\ell -}%
\end{array}%
\right) =\left(
\begin{array}{c}
-\frac{1-ie^{-i\zeta _{\ell }}}{1-ie^{i\zeta _{\ell }}} \\
1%
\end{array}%
\right)%
\end{array}%
\end{equation}

From the above results, and under the symmetric choice of parameters, we can
compute the overlap matrix, $M_{\alpha\beta}=\left\langle
\alpha\right|\beta\rangle$. Neglecting exponentially suppressed corrections,
we obtain the following form:
\begin{equation}
\begin{array}{c}
M_{\alpha\beta}=v\delta_{\alpha\beta}\left[\frac{B+\Delta_m\cos(\phi_{%
\alpha}-\phi_m)}{2(B^2-\Delta_m^2)}\right. \\
\left. +\frac{\Delta_r(B+\Delta_r)+\mu^2}{2\Delta_r(\Delta_r^2-B^2+\mu^2)}%
\right], \label{m}%
\end{array}%
\end{equation}
with $v$ being the spin-orbit velocity.

The coupling between the Majoranas could be calculated perturbatively by
considering the two domain walls juxtaposed. For instance, while the left
Majorana is an exact zero-energy eigenstate of the Hamiltonian
\[
{\mathcal{H}}_{\mathit{L}}={\mathcal{H}}_{\ell}\Theta(-x)+{\mathcal{H}}%
_m\Theta(x),
\]
the existence of the right segment of the wire perturbs this wave function,
with the perturbation potential being
\[
V_r=({\mathcal{H}}_r-{\mathcal{H}}_m)\theta(x-L).
\]
Similarly, we can write ${\mathcal{H}}={\mathcal{H}}_{\mathit{R}}+V_{\ell}$
with $V_{\ell}=({\mathcal{H}}_{\ell}-{\mathcal{H}}_m)\theta(-x)$. This
perturbation induces a hybridization matrix between the left Majorana and
the right Majorana:
\begin{equation}
h=\left(%
\begin{array}{cc}
\left\langle \mathit{L}\right|V_r\left| \mathit{L}\right\rangle=0 &
\left\langle \mathit{L}\right|V_r\left| \mathit{R}\right\rangle \\
\left\langle \mathit{R}\right|V_r\left| \mathit{L} \right\rangle &
\left\langle \mathit{R}\right|V_{\ell}\left| \mathit{R}\right\rangle=0%
\end{array}%
\right).
\end{equation}
In our case,
\[
\begin{array}{c}
V_r=\left[\left(\Delta_r\cos\phi_r-\Delta_m\cos\phi_m\right)\tau^x-\right.
\\
\left. \left(\Delta_r\sin\phi_r-\Delta_m\sin\phi_m\right)\tau^y-\mu_r\tau_z
\right]\Theta(x-L).%
\end{array}%
\]
The perturbation matrix we obtain is:
\begin{equation}
\begin{array}{c}
h=i v e^{i\nu\epsilon_{\alpha\beta}}\epsilon_{\alpha\beta}\left[%
e^{-L/\lambda^m_+}\sin\frac{\phi_r-\phi_m}{2}\sin\frac{\phi_{\ell}-\phi_m}{2}%
\right. \\
\left. +e^{-L/\lambda^m_-}\cos\frac{\phi_r-\phi_m}{2}\cos\frac{%
\phi_{\ell}-\phi_m}{2}\right].%
\end{array}%
\end{equation}
with $\nu$ an unimportant phase.

We arrive at the final answer for the Josephson coupling using Eq. (\ref{tbm}%
). The result indeed coincides with Eq. (\ref{hm}):
\begin{equation}
\begin{array}{c}
{\mathcal{H}}_{JM}=(2 f^{\dagger}f-1)\left[J_+ e^{-L/\lambda_+}\sin\frac{%
\phi_r-\phi_m}{2}\sin\frac{\phi_{\ell}-\phi_m}{2}\right. \\
\left. +J_-e^{-L/\lambda_-}\cos\frac{\phi_r-\phi_m}{2}\cos\frac{\phi_l-\phi_m%
}{2}\right] \\
=(2 f^{\dagger}f-1)\left(J_M\cos\frac{\phi_r-\phi_{\ell}}{2}+J_Z\cos\left(%
\frac{\phi_r+\phi_{\ell}}{2}-\phi_m\right)\right)%
\end{array}%
\end{equation}
with the constants $J_{\pm}$ being:
\begin{equation}
J_+=J_-\approx\frac{v}{\overline{M}_{rr}}.
\end{equation}
where $\overline{M}_{rr}=v\left[\frac{B}{2(B^2-\Delta_m^2)}+\frac{%
\Delta_r(B+\Delta_r)+\mu^2}{2\Delta_r(\Delta_r^2-B^2+\mu^2)}\right].$ is the
average of the overlap matrix [Eq. (\ref{m})] diagonal elements, dropping
the cosine term. The cosine term in the overlap will produce additional
harmonics of the Majorana-Josephson term but will not qualitatively change
the answer we obtained. The $J_{\pm}$ terms give rise to to the previously
explored Majorana-Josephson term, Eq. (\ref{J1}) and to the new zipper term,
Eq. (\ref{J2}).

\section{Numerical Calculation}

We now detail the procedure of our numerical calculation. In the Nambu
spinor basis $\Psi ^{T}=(\psi _{\uparrow },\psi _{\downarrow },\psi
_{\downarrow }^{\dagger },-\psi _{\uparrow }^{\dagger })$, the Bogoliubov-de
Gennes Hamiltonian for this system is
\begin{equation}
{\mathcal{H}}=v\hat{p}\sigma ^{z}\tau ^{z}-\mu \tau ^{z}+\Delta \left( \cos
\phi \tau ^{x}-\sin \phi \tau ^{y}\right) +B\sigma ^{x},
\end{equation}%
with $v$ the edge-state velocity, $\hat{p}$ the momentum, $B$ the Zeeman
energy, and $\sigma ^{a}$ and $\tau ^{a}$ Pauli matrices acting in the spin
and particle-hole sectors, respectively. We allow the chemical potential $%
\mu $, pairing amplitude $\Delta $, and superconducting phase $\phi $, to
vary spatially. In region $s$ (with $s=l,m,r$), the parameters $\left( \mu
,\Delta ,\phi \right) =\left( \mu _{s},\Delta _{s},\phi _{s}\right) $ are
constant. Without loss of generality, we assume $\phi _{m}=0$ to be a
reference of superconducting phase.

The Josephson effects in the TST junction has both bound states and
continuum contributions. In the following, we first present the procedure to
compute the exact interaction energy $E$ between two Majoranas, and then
provide the formalism to calculate the energy contribution from the
continuum.

\subsection{Bound state energy}

For TST configuration, there are two Majoranas at interfaces between
topological and trivial regions. The finite separation leads to a finite
interaction energy $E=E_{int}$ between these two Majoranas, with
spatial-dependent wave function $\Psi =\Psi \left( x\right) $ satisfying the
equation%
\begin{equation}
{\mathcal{H}}\Psi =E\Psi .
\end{equation}%
We will solve the interaction energy $E=E_{int}$ by matching the boundary
condition of the wave function.

First, we replace the momentum operator $\hat{p}$ with $-i\frac{\partial }{%
\partial x}$, and obtain the linear differential equation associated with
energy $E$%
\begin{equation}
\frac{\partial }{\partial x}\Psi \left( x\right) =\mathbf{G}_{E}\Psi \left(
x\right) ,
\end{equation}%
with $4\times 4$ matrix%
\begin{equation}
\mathbf{G}_{E}=i\frac{\mu }{v}\sigma ^{z}+\frac{\Delta }{v}\sigma ^{z}\left(
\cos \phi \tau ^{y}+\sin \phi \tau ^{x}\right) -\frac{B}{v}\sigma ^{y}\tau
^{z}+i\frac{E}{v}\sigma ^{z}\tau ^{z}.
\end{equation}%
In region $s$ (with $s=l,m,r$), the parameters $\left( \mu ,\Delta ,\phi
\right) =\left( \mu _{s},\Delta _{s},\phi _{s}\right) $ are constant, and
the matrix $G_{E}^{\left( s\right) }$ has eigensystem%
\begin{equation}
\mathbf{G}_{E}^{\left( s\right) }\vec{u}_{j}^{\left( s\right) }=\kappa
_{j}^{\left( s\right) }\vec{u}_{j}^{\left( s\right) }
\end{equation}%
with eigenvalues $\kappa _{j}^{\left( s\right) }$ and eigenvectors $\vec{u}%
_{j}^{\left( s\right) }$ for $j=1,\cdots ,4$ and $s=l,m,r$.

Then, we expand the four-component wave function $\Psi \left( x\right) $ in
terms of eigenvectors $\vec{u}_{j}^{\left( s\right) }$. We are interested in
the localized state with $E<E_{gap}^{\left( l,r\right) }$. In the left
region, there are two localized modes ($\mathrm{Re}\kappa _{1,2}^{\left(
l\right) }>0$) and the two divergent modes ($\mathrm{Re}\kappa
_{3,4}^{\left( l\right) }<0$). Similarly, in the right region, there are two
localized modes ($\mathrm{Re}\kappa _{1,2}^{\left( r\right) }<0$) and the
two divergent modes ($\mathrm{Re}\kappa _{3,4}^{\left( r\right) }>0$). The
wave function with two localized Majoranas consists of localized modes%
\begin{equation}
\Psi \left( x\right) =\left\{
\begin{tabular}{ll}
$\sum\limits_{j=1,2}c_{j}^{\left( l\right) }e^{\kappa _{j}^{\left( l\right)
}x}\vec{u}_{j}^{\left( l\right) }$ & for $x\leq 0$ \\
$\sum\limits_{j=1,2}c_{j}^{\left( r\right) }e^{\kappa _{j}^{\left( r\right)
}\left( x-L\right) }\vec{u}_{j}^{\left( r\right) }$ & for $x\geq L$%
\end{tabular}%
\right. .
\end{equation}%
In order to match the coefficients associated with left and right regions,
we integrate the wavefunction over the middle region and obtain the condition%
\begin{equation}
\sum\limits_{j=1,2}c_{j}^{\left( r\right) }\vec{u}_{j}^{\left( r\right)
}=\Psi \left( L\right) =e^{\mathbf{G}_{E}^{\left( m\right) }L}\Psi \left(
0\right) =\sum\limits_{j=1,2}c_{j}^{\left( l\right) }e^{\mathbf{G}%
_{E}^{\left( m\right) }L}\vec{u}_{j}^{\left( l\right) },
\end{equation}%
which can be written as%
\begin{equation}
\mathbf{M}_{E}\left(
\begin{array}{c}
c_{1}^{\left( l\right) } \\
c_{2}^{\left( l\right) } \\
c_{1}^{\left( r\right) } \\
c_{2}^{\left( r\right) }%
\end{array}%
\right) =\left(
\begin{array}{c}
0 \\
0 \\
0 \\
0%
\end{array}%
\right)
\end{equation}%
with $4\times 4$ matrix%
\begin{equation}
\mathbf{M}_{E}=\left[
\begin{array}{cccc}
\left( e^{\mathbf{G}_{E}^{\left( m\right) }L}\vec{u}_{1}^{\left( l\right)
}\right) & \left( e^{\mathbf{G}_{E}^{\left( m\right) }L}\vec{u}_{2}^{\left(
l\right) }\right) & \left( -\vec{u}_{1}^{\left( r\right) }\right) & \left( -%
\vec{u}_{2}^{\left( r\right) }\right)%
\end{array}%
\right] .
\end{equation}

\begin{figure}[tbp]
\centering
\includegraphics[width=8.7cm]{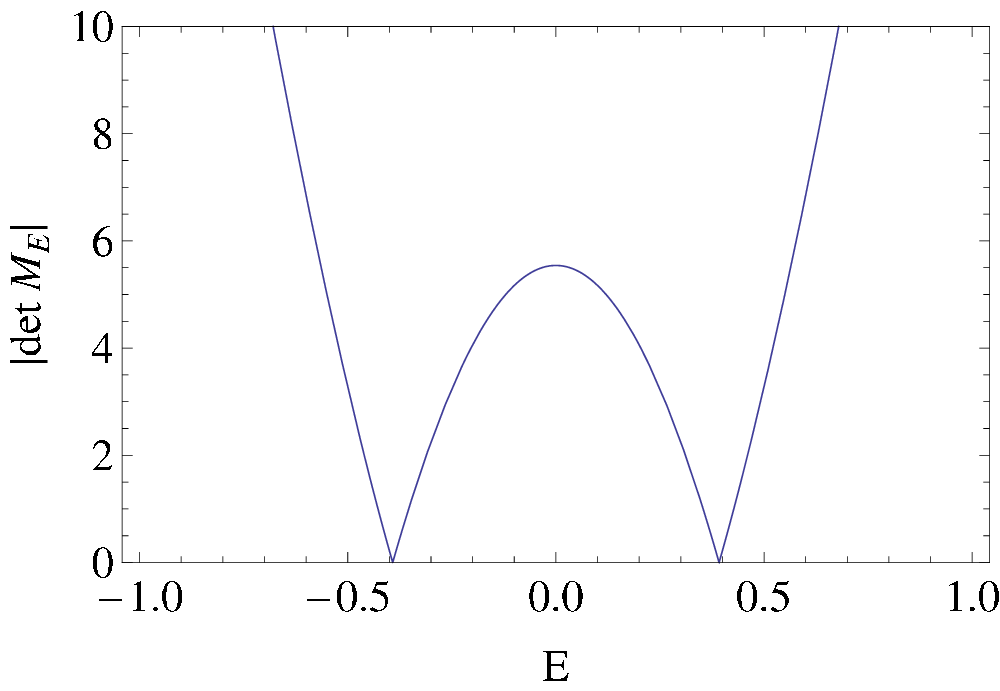}
\caption[fig:Det]{{}The function $\det M_{E}$ vanishes when $E=\pm E_{int}$,
which can be used to numerically find the interaction energy between the
Majoranas.}
\label{fig:Det}
\end{figure}

The necessary condition for non-zero solution is%
\begin{equation}
\det \mathbf{M}_{E}=0,
\end{equation}%
which can be used to numerically determine the interaction energy $E_{int}$.
As illustrated in Fig.~\ref{fig:Det}, the function $\det \mathbf{M}_{E}$
vanishes at $E=\pm E_{int}$. (There is a technical subtlety associated with
the fact that $\mathbf{G}_{E}^{\left( l,r\right) }$ is \emph{not} a
Hermitian matrix. For some fixed values of $E$, the eigenvalues of $\mathbf{G%
}_{E}^{\left( l,r\right) }$ have multiplicity larger than one, and the
eigenvector $\vec{u}_{j}^{\left( l,r\right) }$ might be a zero vector, which
may also lead to spurious solutions with vanishing $\det \mathbf{M}_{E}$.
This issue can be resolved by using a polynomial discriminant to identify
and remove these spurious solutions.)

\subsection{Continuum contribution}

We now consider the energy contribution from the continuum. The continuum
states can be characterized by the scattering matrix $\mathbf{S}_{E}=\mathbf{%
S}_{E}\left( \phi _{l},\phi _{r}\right) $, which can be computed by matching
the boundary conditions for all incoming and outgoing modes. Once we know
the scattering matrix, we can use the Fumi's sum rule to compute the
continuum contribution to the system energy \cite{Mahan00,Akkermans91}%
\begin{equation}
W\left( \phi _{l},\phi _{r}\right) =\int_{E_{gap}}^{\infty }\frac{dE}{2\pi i}%
\ln \left[ \det \left[ \mathbf{S}_{E}\left( \phi _{l},\phi _{r}\right) %
\right] \right] .
\end{equation}

The continuum contribution consists of many Fourier components%
\begin{equation}
W\left( \phi _{l},\phi _{r}\right) =\frac{1}{2}\sum_{n_{l},n_{r}=-\infty
}^{\infty }W_{n_{l},n_{r}}\cos (n_{l}\phi _{\ell }+n_{r}\phi _{r})
\end{equation}%
with $W_{n_{l},n_{r}}=W_{-n_{l},-n_{r}}$. Then conventional Josephson terms
are $J_{L/R}=W_{1,0}$ and $W_{0,1}$, and the even harmonics of the zipper
terms are $J_{Z,2n}=W_{n,n}$. In the following, we provide the formalism to
compute the scattering matrix $S_{E}\left( \phi _{l},\phi _{r}\right) .$

For energy $E>E_{gap}^{\left( l,r\right) }$, there are the propagating modes
($\mathrm{Re}\kappa _{j}^{\left( l,r\right) }=0$), with momentum $%
p_{j}^{\left( l,r\right) }=\mathrm{Im}\kappa _{j}^{\left( l,r\right) }$.
Suppose there are four incoming modes $\left( \vec{u}_{1}^{\left( l\right) },%
\vec{u}_{2}^{\left( l\right) },\vec{u}_{1}^{\left( r\right) },\vec{u}%
_{2}^{\left( r\right) }\right) $ with $p_{1,2}^{\left( l\right) }>0$ and $%
p_{1,2}^{\left( r\right) }<0$, and four outgoing modes $\left( \vec{u}%
_{3}^{\left( l\right) },\vec{u}_{4}^{\left( l\right) },\vec{u}_{3}^{\left(
r\right) },\vec{u}_{4}^{\left( r\right) }\right) $ with $p_{3,4}^{\left(
l\right) }<0$ and $p_{3,4}^{\left( r\right) }>0$. The wave function can be
written as a linear combination of all these modes%
\begin{equation}
\Psi \left( x\right) =\left\{
\begin{tabular}{ll}
$\sum\limits_{j=1,\cdots ,4}c_{j}^{\left( l\right) }e^{\kappa _{j}^{\left(
l\right) }x}\vec{u}_{j}^{\left( l\right) }$ & for $x\leq 0$ \\
$\sum\limits_{j=1,\cdots ,4}c_{j}^{\left( r\right) }e^{\kappa _{j}^{\left(
r\right) }\left( x-L\right) }\vec{u}_{j}^{\left( r\right) }$ & for $x\geq L$%
\end{tabular}%
\right. .
\end{equation}%
In order to match the coefficients associated with left and right regions,
we integrate the wavefunction over the middle region and obtain the
condition $\Psi \left( L\right) =e^{\mathbf{G}_{E}^{\left( m\right) }L}\Psi
\left( 0\right) $.
The relation between the amplitudes of incoming and outgoing modes is
\begin{equation}
\mathbf{M}_{E,in}\left(
\begin{array}{c}
c_{1}^{\left( l\right) } \\
c_{2}^{\left( l\right) } \\
c_{1}^{\left( r\right) } \\
c_{2}^{\left( r\right) }%
\end{array}%
\right) =\mathbf{M}_{E,out}\left(
\begin{array}{c}
c_{3}^{\left( l\right) } \\
c_{4}^{\left( l\right) } \\
c_{3}^{\left( r\right) } \\
c_{4}^{\left( r\right) }%
\end{array}%
\right)
\end{equation}%
with $4\times 4$ matrices%
\begin{equation}
\mathbf{M}_{E,in}=\left[
\begin{array}{cccc}
\left( e^{\mathbf{G}_{E}^{\left( m\right) }L}\vec{u}_{1}^{\left( l\right)
}\right)  & \left( e^{\mathbf{G}_{E}^{\left( m\right) }L}\vec{u}_{2}^{\left(
l\right) }\right)  & \left( -\vec{u}_{1}^{\left( r\right) }\right)  & \left(
-\vec{u}_{2}^{\left( r\right) }\right)
\end{array}%
\right] ,
\end{equation}%
\begin{equation}
\mathbf{M}_{E,out}=\left[
\begin{array}{cccc}
\left( -e^{\mathbf{G}_{E}^{\left( m\right) }L}\vec{u}_{3}^{\left( l\right)
}\right)  & \left( -e^{\mathbf{G}_{E}^{\left( m\right) }L}\vec{u}%
_{4}^{\left( l\right) }\right)  & \left( \vec{u}_{3}^{\left( r\right)
}\right)  & \left( \vec{u}_{4}^{\left( r\right) }\right)
\end{array}%
\right] .
\end{equation}%
The scattering relation is%
\begin{equation}
\mathbf{S}_{E}\left(
\begin{array}{c}
\left( p_{1}^{\left( l\right) }\right) ^{1/2}c_{1}^{\left( l\right) } \\
\left( p_{2}^{\left( l\right) }\right) ^{1/2}c_{2}^{\left( l\right) } \\
\left( p_{1}^{\left( r\right) }\right) ^{1/2}c_{1}^{\left( r\right) } \\
\left( p_{2}^{\left( r\right) }\right) ^{1/2}c_{2}^{\left( r\right) }%
\end{array}%
\right) =\left(
\begin{array}{c}
\left( -p_{3}^{\left( l\right) }\right) ^{1/2}c_{3}^{\left( l\right) } \\
\left( -p_{4}^{\left( l\right) }\right) ^{1/2}c_{4}^{\left( l\right) } \\
\left( -p_{3}^{\left( r\right) }\right) ^{1/2}c_{3}^{\left( r\right) } \\
\left( -p_{4}^{\left( r\right) }\right) ^{1/2}c_{4}^{\left( r\right) }%
\end{array}%
\right)
\end{equation}%
with scattering matrix%
\begin{equation}
\fbox{$\mathbf{S}_{E}=\mathbf{P}_{out}^{1/2}\cdot \mathbf{M}%
_{E,out}^{-1}\cdot \mathbf{M}_{E,in}\cdot \mathbf{P}_{in}^{-1/2},$}
\end{equation}%
where $\mathbf{P}_{in}=\mathrm{Diag}\left[ p_{1}^{\left( l\right)
},p_{2}^{\left( l\right) },p_{1}^{\left( r\right) },p_{2}^{\left( r\right) }%
\right] $ and $\mathbf{P}_{out}=-\mathrm{Diag}\left[ p_{3}^{\left( l\right)
},p_{4}^{\left( l\right) },p_{3}^{\left( r\right) },p_{4}^{\left( r\right) }%
\right] $. The requirement of conservation of current is%
\begin{equation}
\sum_{j}p_{j}^{\left( l\right) }\left\vert c_{j}^{\left( l\right)
}\right\vert ^{2}=\sum_{j}p_{j}^{\left( r\right) }\left\vert c_{j}^{\left(
r\right) }\right\vert ^{2},
\end{equation}%
which ensures the unitarity of the scattering matrix%
\begin{equation}
\mathbf{S}_{E}^{\dag }\mathbf{S}_{E}=I.
\end{equation}%
Hence, $\det \left[ \mathbf{S}_{E}\right] =e^{i2\delta _{E}}$ and $\frac{1}{%
2\pi i}\ln \left[ \det \left[ \mathbf{S}_{E}\left( \phi _{l},\phi
_{r}\right) \right] \right] =\frac{1}{\pi }\delta _{E}\left( \phi _{l},\phi
_{r}\right) $. Numerically, we just need to compute the quantity $\delta
_{E}\left( \phi _{l},\phi _{r}\right) $ and the integral%
\begin{equation}
W\left( \phi _{l},\phi _{r}\right) =\int_{E_{gap}}^{\infty }\frac{dE}{\pi }%
\delta _{E}\left( \phi _{l},\phi _{r}\right) .
\end{equation}%
The continuum contribution $W\left( \phi _{l},\phi _{r}\right) $ has $2\pi $
periodicity in both $\phi _{l}$ and $\phi _{r}$, with Fourier decomposition
of $W\left( \phi _{l},\phi _{r}\right) =\frac{1}{2}\sum_{n_{l},n_{r}=-\infty
}^{\infty }W_{n_{l},n_{r}}\cos (n_{l}\phi _{\ell }+n_{r}\phi _{r})$, with
Fourier coefficients of $W_{n_{l},n_{r}}$. The relevant Fourier components
are $J_{L/R}=W_{1,0}=W_{0,1}$, and $J_{Z,2n}=W_{n,n}$ for $n=1,2,\cdots $.

There is one subtle issue in the computation of the scattering matrix. There
are four propagating modes for $E>\left\vert B^{\left( l,r\right) }+\sqrt{%
\left( \mu ^{\left( l,r\right) }\right) ^{2}+\left( \Delta ^{\left(
l,r\right) }\right) ^{2}}\right\vert $, but there are two propagating modes
and two localized modes for $\left\vert B^{\left( l,r\right) }+\sqrt{\left(
\mu ^{\left( l,r\right) }\right) ^{2}+\left( \Delta ^{\left( l,r\right)
}\right) ^{2}}\right\vert >E>E_{gap}^{\left( l,r\right) }$ $=\left\vert
B^{\left( l,r\right) }-\sqrt{\left( \mu ^{\left( l,r\right) }\right)
^{2}+\left( \Delta ^{\left( l,r\right) }\right) ^{2}}\right\vert $. In the
latter case, we need to compute the effective scattering matrix that are
projected to the subspace spanned by the propagating modes.

\end{document}